\documentclass[12pt]{article}

\usepackage{latexsym}

\newcommand{\const}{\mathop{\rm const}\nolimits}

\newtheorem{de}{Definition}[section]
\newtheorem{lem}{Lemma}[section]
\newtheorem{theor}{Theorem}[section]

\newtheorem{cor}{Corollary}[section]
\newtheorem{no}{Remark}[section]

\title {Invariant Properties of the Ansatz of the Hirota Method
for Quasilinear Parabolic equations}
\author{ K.A. Volosov \\ 
Moscow Institute of Electronics and Mathematics \\
Contsam@dol.ru }

\begin {document}
\maketitle 


\begin{abstract}
We propose a new method based on the invariant properties 
of the ansatz of the Hirota method 
whcih have been discovered recently. 
This method allows one to construct new solutions 
for a certain class the dissipative equations 
classified by degrees of homogeneity.
This algorithm is similar to the method of ``dressing'' 
the solutions of integrable equations. 
A class of new solutions is constructed. 
It is proved that all known exact solutions of
the  FitzHygh--Nagumo--Semenov equation 
can be expressed in terms of solutions of 
the linear parabolic equation. 
This method is compared with the Miura transforms 
in the theory of Kortveg de~Vris equations. 
This method allows on to create a package 
by using the methods of computer algebra.
\end{abstract}

\section {} 
\label{s1}

We consider the quasilinear parabolic equation 
\begin{eqnarray}
&&(1+b_1u+b_2u^2)u_t-(h_1+h_2u+h_3u^2)u_x-g_0(u_x)^2\nonumber\\
&&\qquad -(k_0+k_1u)u_{xx}+ \sum_{i=1}^{4}\varphi_{i}u^{i}=0.
\label{1}
\end{eqnarray}
where $b_j(x,t)$, $h_j(x,t)$, $k_{j-1}(x,t)$, $j=1,2$,
$\varphi_{i}(x,t)$, $i=1,\dots,4$,
$g_0(x,t)$ are smooth functions or constants.

   It is assumed that the coefficients of  
$u_t$ and$ u_{xx}$ do not vanish in the range of the function~$u$.

    We seek a solution in the form of the fraction 
\cite{8} (R.~Hirota):
\begin {equation}
u(x, t) = \frac {G (x, t)} {F (x, t)}.
\label{2}
\end {equation}
After multiplication by $F(x,t)^{4}$, 
for the functions $G(x,t)$ and $F(x,t)$
we receive a homogeneous equation of the fourth order 
of homogeneity with respect to all functions.
        
A special case of Eq.~(\ref{1}) has the form
\begin {equation}
\label{3}
(1+b_1u)u_t-(h_1+h_2u)u_x-k_0u_{xx}+\sum_{i=1}^{3}\varphi_{i}u^{i}=0,
\end{equation}
which, after the substitution of (\ref{2}) and multiplication by
$F(x,t)^{3}$ becomes  an equation of the third order of homogeneity
with respect to the functions $G(x,t)$ and $F(x,t)$.

Thus this equation is of the third order of homogeneity, and
the the components $u^{4}$ or $(u_x)^{2}$
complicate the equation so that it is an equation of the fourth 
order of homogeneity.

Special cases of this equation are widely  known 
(see the extensive bibliography in~\cite{1,2,3,11,12,14}).

If $h_1=0$, $\varphi_i=0$, $i={1,3}$,
then this is the B\"urgers equation in the wave theory~\cite{4}.

If $h_1=0$, $h_2=0$, $\varphi_1=-1$,
$\varphi_2=-\varphi_1$, $\varphi_3=0$, then this is 
the Fisher--Kolmogorov--Petrovskii--Piskynov (FKPP) equation 
(see the bibliography in~\cite{2,3,13});

If $h_1=0$, $h_2=0$, $\varphi_1=-1$,
$\varphi_2=0$, $\varphi_3=-\varphi_1$, then this is 
the FitzHygh--Nagumo--Semenov (FhNS) equation in the wave theory 
(see bibliography in~\cite{1}--\cite{8}), \cite {19}, p.~12), 
and the Allen--Cahn equation (which is close to the FhNS equation)
in the theory of phase transitions~\cite{10}  
and in ecology \cite{14,15}.
Equations in which the coefficients of the Laplace operator 
depend on the function~$u$ 
appear in models of the theory of epidemic distribution~\cite{20}.
Equations in which the coefficients of the derivative $u_t$
depending on~$u_x$ arise, for example, 
in models of processes of ocean water freshening~\cite{21}. 
The initial reasons on a theme of the given operation can be
 detected in operation ~ \cite {22}.

The method proposed for constructing solutions of Eq.~(1.1) 
ideologically goes back to the Hirota method. 
An essential distinguishing feature of the present paper 
is that a new invariance property of solutions of
equations from a given class is discovered (cf. [16, 17]). 
The equations obtained from (1.1) after the substitution of (1.2) 
can be arranged in the order of homogeneity as embedded equations.

 From now on, we use the following notation for the derivatives:
\begin{equation}
\label{4}
\partial_\nu(\cdot)=(\cdot)_\nu,\qquad\nu=x,t
\end{equation}

We choose the solution (1.2) of Eq.~(1.3) in the form 
\begin{eqnarray}
G(x,t)&=&\exp(H(x,t))B(x,t)+Q(x,t), 
\label{5}\\
F(x,t)&=&\Big(1+A(x,t)\exp(H(x,t))\,\Big). 
\label{6}
\end{eqnarray}

All the other representations of the solution obtained 
by the Hirota method can be reduced to this form.
There is only one possibility to extend this notation, 
namely, to add sums of  series in powers of the exponentials
$\exp(iH(x,t))$, $i=2,\dots$. 
Let us substitute (\ref{5})  first into (\ref{2})  and then into
(\ref{3})  and equate the coefficients of equal powers of $\exp(H(x,t))$  
(these powers are $0,1,2,3$)  with zero. 
 
We denote the obtained system of equations by
$(eq0)$, $(eq1)$, $(eq2)$, $(eq3)$.

Equation~$(eq0)$ is the same original equation (\ref{3}) for the
function $Q(x,t)$. Hence $Q(x,t)$ is a solution of this equation.

Let us consider the equation obtained at a power of
$\exp(3H(x,t))$,  namely, Eq.~$(eq3)$:    
\begin {eqnarray}
&& B\Big(B^{2}\varphi_{3}-2k_0(A_{x})^{2}+B(-b_1A_{t}+H_2A_{x})\Big)
\nonumber\\
&&\qquad
+A\Big(B^2(\varphi_1+\varphi_3)+2k_0A_{x}B_{x}
+B(-A_{t}+B_1B_{t}+h_1A_{x}-h_2B_{x}+k_0A_{xx})\Big)\nonumber\\
&&\qquad
+A^{2}(B\varphi_1+B_{t}-h_1B_{x}-K_0B_{xx})=0.
\label{7}
\end{eqnarray}

The special case considered in \cite{3}, p.~51, Section~24, 
and in [2], p.~190,
is the case of $B=(\pm)A$, where the equation has the form
\begin {equation}
\label{8}
-A^{2}B+B^{3}-ABA_t-2BA_x^{2}+AB\
A_{xx}+A^{2}B_t+2AA_xB_x-A^{2}B_{xx}=0.
\end{equation}
and is an identity.

First, we present a lemma in which we prove that, 
in the conventional ansatz form of a solution, 
there is considerable arbitrariness related to 
the ``extra'' function $A(x,t)$, which must be excluded.
Namely, this function must be added to the function $H(x,t)$
according to the formula 
$H(x,t)_{(\mbox{new})}=H(x,t)_{(\mbox{old})}+\exp(\ln(A))$. 
Then the same notation $H(x,t))_{(\mbox{new})}$ 
is preserved for the function $H(x,t)$.

\begin {lem}
\label{lem1} 
Suppose that the functions $B(x,t)$, $A(x,t)$ in {\rm(1.5)}
belong to $C^{2}[R^{1}\otimes R^{1}_{+}]$ and
satisfy Eq.~$(eq3)$, and the function $B(x,t)$ has the form 
\begin {equation}
B(x,t)=M(x,t)A(x,t).
\label{9}
\end{equation}
Then the function $M(x,t)$ satisfies Eq.~{\rm(1.3)}
\begin {equation}
\Big(1+b_1(x,t)M\Big)M_t
-\Big(h_1(x,t)+h_2(x,t)M\Big)M_x-k_0(x,t)M_{xx}
+\sum_{i=1}^3\varphi_i(x,t)M^{i}=0
\label{10}
\end{equation}
for all functions $A(x,t)$.
\end{lem}

{\it Proof}. 
The proof of the lemma is directly obtained by the substitution
of (\ref{9})  into (\ref{7}). 
Equation~(1.10), which follows from Eq.~(1.7), 
coincides with (1.3).
Hence the function $M(x,t)$ is a solution of this equation. Below
$ M(x,t) $ stands for some {\it known} exact solution of Eq.~(1.3).
$\Box$
\bigskip

It follows from Lemma~1.1 that the number of arbitrary functions
in the ansatz can be decreased. Namely, we write $A=\exp(\ln (A))$ 
and add this summand to the function $H(x,t)$. 
In the present paper,  solutions of the form (\ref{2})  
are called {\it solutions of the first, second, etc. mode}
with respect to their complexity  
(the number of terms containing a power of $\exp(iH)$ as a
factor). 
   
\begin {de}
\label{de1}
The set $\Omega$ of solutions of Eq.~{\rm(1.3)} of the form 
\begin {equation}
u(x,t)=\frac{\exp(H(x,t))M(x,t)+Q(x,t)}{1+\exp(H(x,t))},
\label{11}
\end{equation}
where $M(x,t)$, $Q(x,t)$ are some solutions of Eqs.~{\rm (\ref{3})}
and $H(x,t)$ is an arbitrary function, 
will be called SOLUTIONS OF THE FIRST MODE;
solutions of the form 
\begin {equation}
U (x, t) = 
\frac{\exp (H (x, t)) M (x, t) + \exp (2 H (x, t)) R (x, t) + Q (x, t)}
{1 + Z_1 (x, t) \exp (H (x, t)) + \exp (2 H (x, t)) Z_2 (x, t)},
\label{12}
\end{equation}
will be called SOLUTIONS OF THE SECOND MODE; etc.
\end{de}

We have the following theorem.

\begin{theor}
\label{th1}
Suppose that all coefficients in Eq.~{\rm (\ref{3})} are  constants.
The function $ M $ is a solution of the equation
\begin{equation}
\label{13}
(1+b_1 M)M_t -(h_1+h_2 M)M_x
-k_0 M_{xx}+ \sum_{i=1}^3 \varphi_i M^{i} = 0,
\end{equation}
and $Q$ is a solution of the equation 
\begin{equation}
\label{14}
(1+b_1 Q)Q_t -(h_1 +h_2 Q)Q_x -k_0 Q_{xx}+ 
\sum_{i=1}^{3} \varphi_{i} Q^{i} = 0,
\end{equation}
In addition, they are related by the compatibility relations
\begin{eqnarray}
&&\bigg( (M-Q) b_1 k_0 H_{xx}(x,t)\nonumber\\ 
&&\qquad 
+H_{x}(x,t)\Big(M(b_1h_1-h_2)+Q(-b_1h_1+h_2)
+2b_1k_0 (M_{x}-Q_{x})\Big)\nonumber\\
&&\qquad 
+(2+M b_1+Q  b_1)k_0 H_{x}(x,t)^{2}\nonumber\\
&&\qquad
+(M-Q)\bigg(M^{2}b_1\varphi_3-Q^{2}b_1\varphi_3
+M(-\varphi_3+b_1(\varphi_1+\varphi_3))
-Q(-\varphi_3+b_1(\varphi_1+\varphi_3))\nonumber\\
&&\qquad
+b_1(b_1(M_{t}-Q_{t})+h_2(-M_{x}+Q_{x}))\bigg)\bigg)=0,
 \label{15}
\end{eqnarray}
and, together with the function $H(x,t)$ satisfy the equation
\begin{equation}
-(M-Q)b_1 H_{t}(x,t)+H_{x}(x,t)( M h_2-Q h_2
-2k_0 H_{x}(x,t) )+(M-Q)^{2}\varphi_3=0.
\label{16}
\end{equation}
Then the function $ u(x,t) $  is a solution of Eq.~{\rm(\ref{3})}.
\end{theor}

(The arguments of the functions are omitted here for brevity.). 

\begin{no}
\label{no1}
\rm
    The equation (\ref{13}) is the Riccaty  equation 
for the functions $H^{(1,0)}(x,t)$, and equation (\ref{14}) is 
the Hamilton--Jacobi equation   for the function $H(x,t)$.
\end{no}

   If $b_1=0$, then the statement of Theorem~1.1 becomes simpler 
and the function $H(x,t)$ can explicitly be written in terms of the
 functions $M(x,t)$ and $Q(x,t)$.

\begin{theor}
\label{th2}
Let $b_1=0$. 
Suppose that all coefficients in Eq.~{\rm (\ref{3})}
are constants.
The function $M$ is a solution of the equation
\begin{equation}
\label{17}
 M_t -(h_1+h_2 M)M_x -k_0 M_{xx}
+ \sum_{i=1}^3 \varphi_i M^{i} = 0,
\end{equation}
and $Q$ is a solution of the equation 
\begin{equation}
\label{18}
Q_t -(h_1 +h_2 Q)Q_x -k_0 Q_{xx}
+ \sum_{i=1}^{3} \varphi_{i} Q^{i} = 0.
\end{equation}
In addition, they are related by the compatibility relations
\begin{eqnarray}
\label{19}
&&\bigg(\int (M_{t}-Q_{t})\,dx\bigg)
4( h_2-m)-2(M^{2}-Q^{2})( h_2^{2}+12k_0 \varphi_3-h_2 m)\nonumber\\
&&\qquad
-4(M-Q) (h_1 h_2+ 4 k_0 \varphi_2- h_1 m )
+16 k_0 C_0^{\prime }(t)\nonumber\\
&&\qquad 
+4 k_0 (h_2+3 m)(M_{x}-Q_{x})=0,
\end{eqnarray}
where 
$ m=((\pm)\sqrt{(h_2^{2}+8 k_0 \varphi_3 )})$,
and the function  $ H(x,t)$ has the form 
\begin{equation}
 H(x,t)=C_0(t)-\int 
\frac{(M(x,t)-Q(x,t))\Big(-h_2\pm\sqrt{h_2^{2}+8k_0\varphi_3}\Big)}
{4k_0} \,dx.
\label{20}
\end{equation}
Then the function $ u(x,t)$ is a solution of Eq.~{\rm(\ref{3}) }.
\end{theor}

{\it Proof}.
We give a brief proof of this theorem. 
We substitute (\ref{11})  into (\ref{3}),  reduce to the common
denominator, and consider the numerator of the obtained large
equation. 

We equate with zero 
the coefficient of the exponential raised to the zero power, 
i.e., the coefficient of $\exp(0)$. 
The equation thus obtained is exactly Eq.~(\ref{3})  for 
the function $ Q(x,t) $. Hence the function $ Q(x,t) $ is a
solution of this equation.

We equate with zero 
the coefficient of the exponential $\exp (3H (x, t)) $.
The equation thus obtained 
will be called Eq.~$(eq3)$. Again it is exactly  Eq.~(\ref{3}).
Hence the function $M(x, t)$ is a solution of this equation.

We equate with zero 
the coefficient of the exponential $\exp (H (x,t))$.
The equation thus obtained 
will be called Eq.~$(eq1)$.

We equate with zero 
the coefficient of the exponential $\exp(2H(x,t))$.
The equation thus obtained 
will be called Eq.~$(eq2)$. 
Since the functions $Q$ and $M$ satisfy Eq.~(1.3), 
we exclude the second derivatives of the functions $Q_{xx}$
and $M_{xx}$
from Eqs.~$(eq1)$ and~$(eq2)$.

We express $H_{xx}$ from Eq.~$(eq1)$ and stubstitute it into 
Eq.~$(eq2)$. Then we obtain the Hamilton--Jacobi equation (1.16). 
For $b_1=0$ this equation in Theorem~1.2 naturally determines
the function $H(x,t)$ (1.20). 
If we express the derivative $ H_t$ from Eq.~(1.16) and substitute
it into Eq.~$(eq1)$, then we obtain the compatibility condition
(1.15) or (1.19), respectively. 
If we substitute the derivative $ H_t $ into Eq.~$(eq2)$, then we also
obtain the compatibility condition (i.e., the two equations
coincide). 
In conclusion we note that there exist two branches 
in the relations given in the theorem with different signs. 
The proof of the theorem is complete.
$\Box$
\bigskip

In fact, the theorems proved above describe the procedure of
constructing new solutions of the equations 
(an analog of the superposition of solutions), 
which is specified by the properties of Eq.~(\ref{3}) and the
ansatz. 

Let us consider two smooth solutions of Eq.~(\ref{3}):
$M (x, t) $ and $Q (x, t) $. The following function is assign
to these solutions: 
\begin {equation}
P(Q,M)=\frac{Q(x,t)+M(x,t)\exp(H(x,t))}{1+\exp(H(x,t))},
\label{21}
\end{equation}
where $H(x,t)$ is determined by (1.16) or by formula (1.20) if
$b_1=0$. 

The following theorem states that the solutions are invariant. 

\begin{theor}
\label{th3}
Suppose that two pairs of functions
$$
\Big(M_0(x,t),Q_0(x,t)\Big)\qquad \mbox{and}\qquad
\Big(M_1(x,t),Q_1 (x,t)\Big)
$$
are solutions of the equations    
\begin{equation}
\label{22}
 (1+b_1 M_j){M_j}_{t} -(h_1+h_2 M_j){M_j}_{x}
 -k_0 {M_j}_{xx}+ \sum_{i=1}^3 \varphi_i {M_j}^{i} = 0,
\end{equation}
\begin{equation}    
\label{23}
 (1+b_1 Q_j){Q_j}_{t} -(h_1 +h_2 Q_j){Q_j}_{x}
 -k_0 {Q_j}_{xx}+ \sum_{i=1}^{3} \varphi_{i}{Q_j}^{i} = 0,
\end{equation}
where $j=0,1$, and that 
they determine the solution by formula {\rm(1.11)}.
Suppose that the assumptions of Theorems~{\rm1.1} and~{\rm1.2}
hold  for each pair.
Then the two functions
\begin{eqnarray}
Q(x,t)&=& \frac{\exp(H_1 (x,t))M_1 (x,t)+Q_1 (x,t)}
 {1+\exp(H_1 (x,t))},\nonumber\\
M(x,t)&=& \frac{\exp(H_0 (x,t))M_0 (x,t)+Q_0 (x,t)}
 {1+\exp(H_0 (x,t))}
 \label{24}
\end{eqnarray}
are solutions of the equations    
\begin{equation}
\label{25}
(1+b_1 M)M_t -(h_1+h_2 M)M_x -k_0 M_{xx}+ \sum_{i=1}^3 \varphi_i M^{i} = 0,
\end{equation}
\begin{equation}
\label{26}
(1+b_1 Q)Q_t -(h_1 +h_2 Q)Q_x -k_0 Q_{xx}+ \sum_{i=1}^{3}\varphi_{i} Q^{i}=0,
\end{equation}
and the function
\begin{equation}
 u(x,t)=\frac{\exp(H(x,t))M(x,t)+Q(x,t)}{1+\exp(H(x,t))}
\label{27}
\end{equation}
is a solution of Eq.~{\rm(\ref{3})}.
\end{theor}
  
{\it Proof}.
Each of the functions $M_j$, $Q_j$,  $j=0,1,$, 
satisfies Eq.~(\ref{3}) written for this
function. Pariwise, they  satisfy the corresponding
compatibility condition (\ref{15})  or~(\ref{19}). 
The theorem can be proved by a direct substitution.
$\Box$
\bigskip

\begin{no}
\label{no2}
\rm
After calculating the derivative with respect to~$x$,
it is possible to consider 
the compatibility condition~(\ref{15})
as a modified  B\"urgers equation with a potential.    
\end{no}
  
\begin{cor}
\label{cor1}
The class of solutions {\rm(\ref{27})} is closed with respect to
the procedure described in Theorem~{\rm 1.3}. 
Thus {\rm(\ref{27})} defines a surjective mapping:
\begin {equation}
\mbox{Vol}:\,\Omega\mapsto\Omega.
\label{28}
\end {equation}
\end {cor}
 
Similar results take place for Eq.~(1.1). 
We perform the change of the variables 
$\varphi_{1}=a_{2}a_{1}$,
$\varphi_{2}=-(a_{1}+a_{2}+a_{1}a_{2})$,
$\varphi_{3}=1+a_{1}+a_{2}$, $\varphi_{4}=-1$,
which is equivalent to a source-sink function written so that  
the four roots are explicitly distinguished: 
$u(a_1-u)(a_2-u)(1-u)$. 

We have a lemma similar to Lemma~1.1 and the following theorem.
 
\begin {theor}
\label{th4}
Suppose that the solution {\rm(1.1)} has the form
{\rm(1.11)} and the relations hold in one of the following
cases{\rm:} 

{\rm1)} $(Q(x,t)$ or $M(x,t))\equiv\const=-k_0/k_1$,
$(k_0+k_1)(k_0+a_1k_1)(k_0+2a_1k_1)=0$;

{\rm2)} $b_1=k_1/k_0$, $b_2=0$;

{\rm3)} $b_1=0$, $b_2=0$, $k_1=0$.

Then the functions $M$ and $Q$ are also solutions of
an equation coinciding with Eq.~{\rm(1.1)}. 
Moreover, these functions are related by the  
compatibility relations 
{\rm(}the Riccati equation for $H_{x}(x,t)${\rm)}. 
In addition, the Hamilton--Jacobi equation must be satisfied.  
\end{theor}

(Here we do not write these equations, since they are very
cumbersome.) The proof is similar to the proofs of the preceding
theorems.

\section{}
\label{s2}

In the case of variable coefficients, there are additional
possibilities to satisfy the system of equations. Since we
construct exact solutions, not all coefficients of the equation
are arbitrary. 

We have the following theorem.
   
\begin{theor}
\label{th5}
Suppose that $b_1(x,t)=0$,
the function $M(x,t)$ is a solution of the equation 
\begin{equation}
\label{29}
M_t-\Big(h_1(x,t)+h_2(x,t) M\Big)M_x
-k_0(x,t) M_{xx}+ \sum_{i=1}^3 \varphi_i(x,t) M^{i} = 0,
\end{equation}
the function $Q(x,t)$ satisfies the modified equation 
\begin{equation}
\label{30}
Q_t-\Big(h_{01}(x,t)+h_{02}(x,t) Q\Big)Q_x
-k_{00}(x,t) Q_{xx}+ \sum_{i=4}^{6} \varphi_{i}(x,t) Q^{i-3} = 0,
\end{equation}
and the coefficients in the equations satisfy the relations
\begin{eqnarray}
&& Q\Big( \varphi_1(x,t)-\varphi_4(x,t)
+Q(\varphi_2(x,t)-\varphi_5(x,t))
+Q^{2}(\varphi_3(x,t)-\varphi_6(x,t))\Big)\nonumber\\
&&
+\Big(h_{01}(x,t)-h_{1}(x,t)
+(h_{02}(x,t)-h_{2}(x,t))Q\Big)Q_x \nonumber\\
&&-( k_0 Q_{xx}- k_{00}Q_{xx})=0,
\label{31}
\end{eqnarray}
and the functions $M$, $Q$ and the coefficients satisfy the
second compatibility equation 
\begin{eqnarray}
\label{32}
&&\bigg(\int\partial_t\bigg(\frac{(-M+Q)(h_2\pm m)}{k_0}\bigg)\,dx\bigg)
4k_0 (-M+Q)m\nonumber\\
&&+2(M-Q)(\pm h_2^{3}(M^{2}-Q^{2})
+2h_1(M-Q)((\pm)m^{2}+h_2 m)\nonumber\\
&&+ {h_2}^{2}\Big( (M-Q)(M+Q)m-(\pm)2 M {k_0}_{x}
+ (\pm)2Q {k_0}_{x}+ (\pm)6 k_0 (M_{x} -Q_{x})\Big) \nonumber\\
&&+2h_2\Big((\pm)4 k_0 \varphi_3 (M^{2}-Q^{2})
+(\pm)k_0 {h_2}_{x}(M-Q)
-m({k_0}_{x}(M-Q)+k_0(M_{x}-Q_{x})\Big)\nonumber\\
&&+2k_0\bigg(m\bigg(2(M-Q)(2\varphi_2+3\varphi_3(M+Q))-
4C_0^{\prime}(t)+{h_2}_{x}(M-Q)\nonumber\\
&&- (\pm)4 \varphi_3 {k_0}_{x}(M-Q)+  (\pm)24\varphi_3 k_0(M_{x}-Q_{x})
+(\pm)4 {\varphi_3}_{x} k_0(M-Q)\bigg)\bigg) =0,
\end{eqnarray}
where 
$m=(\sqrt{(h_2^{2}+8 k_0 \varphi_3 )})$ 
and the function $H(x,t)$ has the form 
\begin{equation}    
H(x,t)=C_0(t)-\int\frac{(M(x,t)-Q(x,t))
\Big(-h_2 (x,t)\pm\sqrt{h_2 (x,t)^{2}+8k_0 (x,t)\varphi_3 (x,t)}\Big)}
{4k_0 (x,t)} \,dx.
\label{33}
\end{equation}
Then the function $u(x,t)$ {\rm(\ref{11})} is a solution of
Eq.~{\rm(\ref{3})}. 
\end{theor}

{\it Proof}.
The proof is similar to that of Theorem~1.2. Note that, 
first, in this case the set of solutions to the system of
equations for the functions $M$, $Q$ and the coefficients 
of the equation described in the theorem is not empty;

second, it is possible to construct solutions with variable
roots of the source-sink function.

We substitute (1.11) into (1.3),  reduce to the common
denominator, and consider the numerator of the obtained 
equation. 
By equating with zero the coefficient of the exponential raised
to the zero power,  we obtain Eq.~$(eq0)$, which is the
compatibility condition for coefficients~(2.22). 
From this condition we express some coefficient and exclude this
coefficient from the other equations. 
The equation obtained by equating 
the coefficient of the exponential $ \exp (H (x, t)) $ with zero
is called $(eq1)$. 
The equation obtained by equating 
the coefficient of the exponential $ \exp (2H (x, t)) $ with
zero is called $(eq2)$. 
Since the function~$Q$ satisfies Eq.~(2.30), 
we exclude the derivative $Q_t(x, t)$ 
from Eqs.~$(eq1)$ and $(eq2)$.
We can express $H_{xx}(x, t)$ from Eq.~$(eq1)$ and substitute it
into Eq.~$(eq2)$. Then we obtain the Hamilton--Jacobi equation,
which naturally determines the function $H(x,t)$ (2.24). 
The proof of the theorem is complete. 

In the construction of the solution of the second mode, 
the transformation  
relates the solution of the original equation 
and the solutions of two modified equations, and so on.
   
{\it The  FitzHygh--Nagumo--Semenov  equation}

{\it EXAMPLE 1}.
We apply our method to the FitzHygh--Nagumo--Semenov (FhNS)
equation known in wave theory: 
\begin {equation}
\label{34}
u_t-\varphi_{3}u_{xx}/(2a^{2})-\varphi_{3}u+\varphi_{3}u^{3}.
\end{equation}
where $k_0=\varphi_{3}/(2a^2)$ and 
$\varphi_{1}=\varphi_{3}$   are constants.
We show that this method leads to new results and provides a new
understanding of the results (see [3], p.~64, and [9]).

First we show the action of the semigroup related to the
transformation (1.11) and explain how the translation constants
pass into new solutions,
by choosing the form of the functions
(of its solutions, [1], p.~17, [3], p.~51)
satisfying the assumptions of Theorem~1.2.
Namely, we consider the functions
$u(x,t)=M(x,t)$ or $u(x,t)=Q(x,t)$:
\begin {eqnarray}
\label{35}
M (x, t) & = & - (1 + \exp (m_0 + a x + b t)) ^ {-1}, \\
Q (x, t) & = & (-1 + \exp (2a x + q_0)) / (1 + \exp (2a x + q_0)),
\qquad b = -3\varphi_3/2. \nonumber 
\end {eqnarray}
We calculate the function
\begin {equation}
H (x, t) = \ln\frac{1 + \exp (m_0 + a x-3 t\varphi_3 /2)}
{1 + \exp (q_0 + 2ax)},
\label{36}
\end {equation}
and obtain the solution
\begin {equation}
u_1 = \frac{1-\exp (-q_0-2a x + \ln (2))}
{1 + \exp(-q_0-2a x + \ln (2)) + \exp(m_0-q_0-a x-3t\varphi_3/2)}.
\label{37}
\end {equation}
Thus the original transition constants $q_0$, $m_0$ pass into a
new solution and a new constant $\ln(2)$ appears.

Note that this is a well-known solution that describes the wave
interaction ([2], p.~190, [3], p.~51, and [7]).
Below we prove that this solution can be expressed via
the solution of the linear parabolic equation.

There exists another version of the functions for which the
assumptions of Theorem~1.2 are satisfied:
\begin {eqnarray}
\label{38}
M (x, t) & = & (1 + \exp (m_1-a x + b t)) ^ {-1}, \\
Q (x, t) & = & (1-\exp (q_1-2a x)) / (1 + \exp (q_1-2a x)), \qquad 
b = -3\varphi_3/2.\nonumber
\end {eqnarray}
Then the function $H(x,t)$ has the form
\begin {equation}
H (x, t) = \ln\frac{1 + \exp(m_1-a x-3 t\varphi_3 /2}{1+\exp (q_1-2ax)},
\label{39}
\end {equation}
and we obtain the solution 
\begin {equation}
u_2 = \frac{1-\exp (-q_1 + 2a x + \ln (2))}
{1 + \exp(-q_1 + 2a x + \ln (2)) + \exp(m_1-q_1 + a x-3 t\varphi_3/2)}.
\label{40}
\end {equation}
For the new functions $Q(x,t)=u_1$, $M(x,t)=u_2$, 
we take the solutions of Eq.~(2.34):
\begin{eqnarray}
H (x, t)& = &\ln\bigg(
\frac{2\exp (-q_1) + \exp(-2a x) + \exp(m_1-q_1-a x-3 t\varphi_3/2)}
{1 + 2 \exp (q_0-2a x)} \nonumber\\
&&\quad +\exp (m_0-q_0-a x-3 t\varphi_3/2)\bigg), \nonumber \\
u_3 & = &
\bigg(-1+\exp\bigg(q_0-q_1+2ax
+\ln\frac{2+\exp(q_1)}{2+\exp(q_0)}\bigg)\bigg)\nonumber\\
&&\quad\times
\bigg(
1+\exp\bigg(q_0-q_1+2ax+\ln\frac{2+\exp(q_1)}{2+\exp(q_0)}\bigg)\nonumber\\
&&\quad 
+\exp\bigg(-q_1+ax
+\ln\frac{\exp(m_1+q_0)+\exp(m_0+q_1)}{2+\exp(q_0)}
-3\frac{ t\varphi_3}2\bigg)\bigg)^{-1}.
\label{41}
\end{eqnarray}
By analyzing this formula, it is possible to understand 
the structure of the translation constant. Indeed, 
in the variables $\tau_1=-ax+3t\varphi_3/2$,
$\tau_2=ax+3t\varphi_3/2$, the solution (2.41) has the form 
\begin{eqnarray}
u_3 (\tau_1, \tau_2) 
&=& \frac{- c_1 \exp (\tau_1)+ c_2 \exp (\tau_2)}
{1 + c_1 \exp (\tau_1) + c_2 \exp (\tau_2)}, \nonumber \\
c_1 
&=&\frac{\exp (q_1) (2 + \exp (q_0))}
{\exp (q_0 + m_1) + \exp (q_1 + m_0)}, \label{42} \\
c_2 
&=&\frac{\exp (q_0) (2 + \exp (q_1))}
{\exp (q_0 + m_1) +  \exp (q_1 + m_0)}.
\nonumber
\end {eqnarray}

\begin {lem}
\label{lem2}
The set $u(x,t)$, $M(x,t)$, $Q(x,t)$ of solutions of the 
FhNS equation {\rm(2.34)} of the form {\rm(2.38), (2.42)}
forms a semigroup with operation $\diamond $ {\rm(1.11)} and the
following properties{\rm:} 

{\rm(a)} commutativity, 
$M(x,t)\diamond Q(x,t)=Q(x,t)\diamond M(x,t)${\rm;}

{\rm(b)} associativity, 
$u(x,t)\diamond (M(x,t)\diamond Q(x,t))=
(u(x,t)\diamond Q(x,t))\diamond M(x,t)${\rm;}

{\rm(c)} any element is the unity for itself, 
$M(x,t)\diamond M(x,t)=M(x,t)$.
\end{lem}

{\it Proof}.
Let us consider two solutions with different sets of constants
$C_i$, $V_i$, $i=1,2$
\begin{eqnarray}
Q(x,t) &=& \frac{-\exp(\tau_1+C_1)+\exp(\tau_2+C_2)}
{1+\exp(\tau_1+C_1)+\exp(\tau_2+C_2)},
\qquad\tau_i=\tau_i[x,t], 
\nonumber \\
M (x, t) &=& 
\frac{-\exp(\tau_1+V_1)+\exp(\tau_2+V_2)}
{1+\exp(\tau_1+V_1)+\exp(\tau_2+V_2)},
\qquad\tau_i=\tau_i(x,t).
\label{43}
\end{eqnarray}
By using Theorem 1.2, we find the solution 
\begin {eqnarray}
u(x,t)&=&M(x,t)\diamond Q(x,t) \nonumber\\
&=&\frac{-\exp(\tau_1)(\exp(C_1)+\exp(V_1))
+\exp(\tau_2)(\exp(C_2)+\exp(V_2))}
{1+\exp(\tau_1)(\exp(C_1)+\exp(V_1))
+\exp(\tau_2)(\exp(C_2)+\exp(V_2))},\nonumber\\
\tau_i&=&\tau_i[x,t],
\label{44}
\end{eqnarray}
which implies the properties given in Lemma~2.1.

Note that the action of the semigroup is a nonlinear
``time-translation'' of the solution. 

As was noted in [2], p.~192, [3], p.~64, [9], and [12],
it is known that Eq.~(2.34) has solutions with singularities 
(monsters, contrast structures) 
whose role cannot be explained. 

   We have the following theorem.

\begin{theor}   
\label{th6}
Suppose that the function $Q(x,t)$ is a solution of the equation
\begin{equation}
\label{46}
Q_t  -k_0 Q_{xx}- \varphi_{1}Q +\varphi_{3}Q^{3}=0,
\end{equation}
the function $U(x,t)$ is a solution  of the 
{\it linear parabolic equation} 
\begin{equation}
U_{t}+\varphi_3 a_1/ a^2 U_{x}-\varphi_3 / a^2 U_{xx}=0,  
\label{47}
\end{equation}
and the compatibility condition 
\begin{equation}
Q_{x}+ Q (a_1-(\pm)a Q-U_{x}/U)=0
\label{48}
\end{equation}
is satisfied.
Then the function $ u(x,t) $  is a solution of Eq.~{\rm(\ref{34})}
and has the form 
\begin{equation}
u=\frac{Q(x,t)U(x,t)\exp(t\varphi_3(1+a_1^2/(2 a^2) ))}
{Q(x,t)\exp(a_1 x )-U(x,t)\exp(t\varphi_3(1+a_1^2/(2 a^2) ))}.
\label{49}
\end{equation}
\end{theor}

{\it  Proof}.
To prove this theorem, we substitute the ansatz 
of the solution of the third mode
\begin{equation}
u (x, t) = \frac{\exp(H(x,t))M(x,t)+\exp(2H(x,t))R(x,t)-Q(x,t)}
{1+Z(x,t)\exp(H(x,t))+\exp(2H(x,t))Z_1(x,t)+\exp(3H(x,t))R(x,t)},
\label{50}
\end{equation}
into Eq.~(2.34), collect similar terms with equal powers of the
exponentials, equate them with zero, and obtain a system of
nine equations. As previously, from these equations  
we successively exclude the second derivatives 
of all functions contained there. 
Since the ansatz is invariant, there is a remarkable fact common
for the algorithm of constructing the solution. 
Namely, just as in the theorems presented above,
the equation becomes factored at some step.
This equaton contains various hints, which allow us to correct 
the ansatz (2.50) and to pass to the next iteration.

Equation $(eq8)$ has the form
\begin {equation}
2a^2\varphi_3+2a^2H_{t}+\varphi_3{H_{x}}^2-\varphi_3H_{xx}=0.
\label{51}
\end{equation}
Its solution has the the form 
\begin{equation}
H(x,t)=-\ln(U(x,t))+a_1x-(2a^2+a_1^2)t\varphi_3/(2a^2),
\label{52}
\end{equation}
and thus implies (2.47).

Equation $(eq0)$ is exactly Eq.~(2.34) for the function $Q(x,t)$.
Hence the function $Q(x,t)$ is a solution of this equation. 

Equation $(eq1)$ at the second iteration step implies the
compatibility conditions.  

The second iteration of Eq.~(eq1) 
implies the compatibility condition in the form of 
the Riccaty equation, in which we study only the situation
corresponding to the sign ``plus''.

Further, at each successive iteration, we have some versions
dur to the fact that there situations not considered earlier. 
We choose 
\begin{eqnarray}
Z(x,t)&=&-(M(x,t)+Q(x,t)^2)/Q(x,t),\nonumber\\
M(x,t)&=&-Q(x,t)^2/2,\nonumber\\
R(x,t)&=&-(2Q(x,t)Z_1(x,t)+Q(x,t)^3)/2.
\label{53}
\end{eqnarray}
By substituting all these expressions into the ansatz (2.50), we
obtain the desired solution. The theorem is proved. 

There is the following analogy with the theory of 
Korteweg-de Vries equtaions. 
In this context, relations (2.48) can be considered as 
an analog of Miura transformation.

Let us prove that all known solutions of the FhNS equation can
be calculated in terms of the solutions of the a linear
parabolic equation. We find the solution of the Riccary equation
(2.48) and, choosing the upper sign, obtain
\begin{equation}
Q(x,t)=\frac{U(x,t)\exp(-a_1x)}{C_0-\int(aU(x,t)\exp(-a_1x)\,dx}.
\label{54}
\end{equation}
The compatibility condtions impose strong restrictions. 
Hence, for example, we have the pair of functions
\begin {eqnarray}
Q(x,t)&=&\frac{1+C_1\exp(2ax)}{1-C_1\exp(2ax)},\nonumber\\
U(x,t)&=&C_1+\exp(-2ax),\qquad a_1=-a.
\label{55}
\end{eqnarray}
the substitution of which into (2.49) implies the following
solution of Eq.~(2.34):
\begin{equation}
u(x,t)=\frac{1+C_1\exp(2ax)}{-1+C_1\exp(2ax)+\exp(ax-3t\varphi_3/2)}.
\label{56}
\end{equation}
And the solution of (2.47)
\begin{eqnarray}
U(x,t)&=&\Big(1+2aC_2\exp(-2ax)\Big)
\exp\Big((a+a1)x+(a-a1)(a+a1)t\varphi_3/(2a^2)\Big),\nonumber\\
a_1&=&Ia\sqrt{2}.
\label{57}
\end{eqnarray}
gives solution of the equation (\ref{34}) 
\begin {equation}
u(x,t)=\frac{2aC_2+\exp(2ax)}{2aC_2-\exp(2ax)+\exp(ax-3t\varphi_3/2)}.
\label{58}
\end{equation}
These solutions have a specific characteristic feature, which
can either appear or disappear. Such solutions were considered
in [3], p. 64, [9], [12].
In what follows, we consider Example~3 constructed by this method.
In this example, {\em two singularities} simultaneously appear
in the modified FhNS equation. 
 
There is the following statement: to the WELL-KNOWN SOLUTION of 
Eq.~(2.34)
\begin{equation}
u=\frac{1-\exp(2ax)}{1+\exp(2ax)+\exp(ax-3t\varphi_3/2)},
\label{59}
\end{equation} 
there corresponds the solution of the 
LINEAR PARABOLIC EQUATION~(\ref{47}) 
\begin {eqnarray}
U(x,t) &=& (-1+\exp(2ax))\exp((-a+a1)x
+(a-a1)(a+a1)t\varphi_3/(2a^2)),\nonumber\\
a_1 &=& Ia\sqrt{2}.
\label{60}
\end{eqnarray}
It turns out that the complex solutions of the linear parabolic
equation can be recalculated into the solutions of the FhNS
equation and vice versa.
Therefore, the complex solutions of these equations must be
studied. 

Let us consider the FitzHygh--Naguma--Semenov (FhNS) equation
(2.34) with variable coefficients.

By Theorem~2.1, for the first mode and separately for the third
mode, there exist the following solutions.
We have the following theorem.

We have the following theorem.

\begin{theor}
\label{th7}
Let $k_0=k_0(x,t)$, $\varphi_{1}=\varphi_{1}(x,t)$,
$\varphi_{3}=\varphi_{3}(x,t)$ be smooth functions.
Suppose that the function $M(x,t)$ is a solution of the equation
\begin{equation}
\label{61}
M_t-k_0M_{xx}-\varphi_{1}M+\varphi_{3}M^{3}=0,
\end{equation}
and the following compatibility condition imposed 
on its coefficients is satisfied:
\begin{eqnarray}
&&-\sqrt{2}\bigg(\int \frac{-M\varphi_{3}{k_0}_{t}
+k_0(2\varphi_{3}M_{t}+{\varphi_{3}}_{t}M)}
{k_0\sqrt{m_2}}\,dx\bigg)\sqrt{m_2}
+ 6 M^2 \varphi_{3} \sqrt{m_2}-4 C_0'(t) \sqrt{m_2}\nonumber\\
&&\qquad 
+ 6\sqrt{2} M_{x}k_0 \varphi_{3}-\sqrt{2} M{k_0}_{x}\varphi_{3}
+ \sqrt{2} M k_0 {\varphi_{3}}_{x}=0,
\label{62}
\end{eqnarray}
where $m_2=k_0(x,t)\varphi_{3}(x,t)$
Then the function 
\begin{equation}
u(x,t)=\frac{ \exp(f)M(x,t)}{1-\exp(f)}, \qquad 
f=C_0(t)+\int \frac{M(x,t)\sqrt{\varphi_{3}(x,t)}}
{\sqrt{2 k_0(x,t)}}\,dx.
\label{63}
\end{equation}
is a solution of the FhNS equation {\rm(2.34)}.
\end{theor}

{\it Proof}.
The proof follows from Theorems 1.2 and 2.1 and can be carried 
out by a direct substitution of the ansatz of the first mode,
taking into account the fact that the coefficients in the
equation are variable and $M=0$.
One compatibility condition is identically satisfied.
One compatibility condition remains.
Hence the set of solutions is not empty. 
The following solution illustrates this theorem.

For example, we set $k_0=1$, $\varphi_{1}=\varphi_{3}$.
We also assume, that we want to dress the solution $Q=1$. 
The compatibility conditon (2.62) turns into the equation for
the function $\varphi_1$:
\begin {eqnarray}
{\varphi_{1}}_{t}-3\sqrt{2\varphi_1}{\varphi_{1}}_{x}+
({\varphi_{1}}_{x})^2/\varphi_1+{\varphi_1}_{xx}=0.  
\label{64}
\end{eqnarray}

We choose one of its solutions, for example,
$\varphi_1={C_2\tanh^{2}(x/2\sqrt{3C_2\sqrt{2}}})/{3\sqrt{2}}$.

The solution of the FhNS equation 
\begin{equation}
\label{65}
u_t-u_{xx}-\varphi_{1}(u-u^{3})=0,
\end{equation}
has the form
\begin{equation}
u(x,t)=\frac{f_1}{1-f_1},\qquad
f_1=\exp\bigg(\frac{C_2t}{2\sqrt{2}}\bigg)
\cosh^{1/3}\bigg(\frac{\sqrt{3C_2}x}{2^{3/4}}\bigg).
\label{66}
\end{equation}
This solution is real. If the constant is $C_2>0$,
then this solution
describes wave interaction . 
At the initial time moment, the solution has a singularity. 
Then the solution becomes smooth. 
For negative values of this constant
$C_2<0 $, this solution describes
a periodic linear structure .

Thus one can construct a solution of the equation with variable
roots of the source-sink function. For example, if $\varphi_1$
is given to be a periodic function, then $\varphi_3$ is
constructed from solutions of Mathieu  and Hill type equations.

\begin {theor}
\label{th8}
Let $k_0=k_0(x,t)$, $\varphi_{1}=\varphi_{1}(x,t)$,
$\varphi_{3}=\varphi_{3}(x,t)$ be smooth functions.
Suppose that one of the compatibility conditions on the
coefficients is satisfied, namely, the first compatibility
equation 
\begin{equation}
Q_{x}(x,t)=\frac{-2\sqrt(2\varphi_{3}k_0)(U Q)^{2} 
-4 k_0 U Q( \varphi_{1}U-U_{x})}{4 k_0 U^{2}},
\label{67}
\end{equation}
or the second compatibility condition  
\begin{equation}
Q_{x}(x,t)=\frac{-2\sqrt{k_0}UQ-2UQ^2\varphi_{3}+2\sqrt{k_0}QU_{x}}
{\sqrt{2k_0}U}
\label{68}
\end{equation}
Suppose that the function $Q(x,t)$ is a solution of the FhNS
equation 
\begin{equation}
\label{69}
Q_t-k_0Q_{xx}-\varphi_{1}Q+\varphi_{3}Q^{3}=0,
\end{equation}
and the function $U(x,t)$
is a solution {\it of the linear parabolic equation }
\begin{equation}
U_{t}+2 \varphi_1 k_0 U_{x}-k_0 U_{xx}
-U\bigg(\int {\varphi_{1}}_{t}\,dx+\varphi_{1}
+k_0\varphi_{1}^2-k_0{\varphi_{1}}_{x}\bigg)=0.  
\label{70}
\end{equation}
Then the function
\begin{equation}
u(x,t)=\frac{Q(x,t)U(x,t)}{Q(x,t)\exp(\int\varphi_{1}\,dx)-U(x,t)},
\label{71}
\end{equation}
is a solution of the FhNS equation {\rm(\ref{34})}.
\end{theor}

{\it Proof}.
This statement can be proved by the substitution of 
the ansatz of the solution of the third mode
(\ref{50}), just as in the
proof of Theorem~2.2. This theorem is a partial case in the
analysis of the one of the two branches, which are determined by
the choice of the compatibility condition.

The method has the following advantages:

a) the algorithm is iterative;

b) at each iteration step we solve one ordinary differential
equation of the first order or one algebraic equation 
(cf. [17], where one has to solve a system of ordinary
differential equations);

c) the method allows one to obtain new solutions; it is
necessary only to know which mode has nontrivial solutions;
to this end, there are some considerations the discussion of
which is beyond the framework of this paper.

Note that formulas (2.41), (2.42) can be also obtained in a
different way, while formulas (2.49), (2.63), (2.72) are
constructed by this method. 

{\it EXAMPLE 2}.
Let us consider a modified FitzHygh--Nagumo--Semenov equation,
which is a special case of the equation (\ref{1}). 
In this case there also exist a lotof solutions, but we write
only one solution correspondign to the second mode, 
since it differs from all other solutions.
The change (\ref{2}) reduces our equation to an equation
the fourth order of homogeneity.

Suppose that Eq.~(\ref{1}) has the form
\begin{equation}
(a_1-h_1u)u_t-(a_1-h_1u)u_{xx}-
h_1(a_1+h_1u)u_x-2h_1(u_x)^{2}-
a_1(a_1^{2}+h_1^{2})u(1-u^{2})/2=0,
\label{72}
\end{equation}
The solution has the form
\begin{equation}
u(x,t)=\frac{1-\exp(a_1x)}
{1+\exp(a_1x)+\exp(((-h_1^{2}-3a_1^{2})t+2(h_1+a_1)x)/4)}
\label{73}
\end{equation}
and describes the  development of the second wave from small
perturbations.

This solution can be used for modeling phenomena related to
phase transitions~\cite{10}. 
One of the elementary equations for which
this effect is preserved for $a_1=0$  has the form 
\begin{equation}
u_t-u_xx+h_1u_x+2(u_x)^{2}/u=0,
\label{74}
\end{equation}

{\it EXAMPLE 3}. 
We consider the  modified FhNS equation (\ref{3}):  
\begin {equation}
u_t+2(h_1)^{2}u_{xx}/\varphi_3-H_1(1+4u)u_x-\varphi_3u(1-u^{2})=0,
\label{75}
\end{equation}

The solution has the form
\begin{equation}
u(x,t)=\frac{1-\exp((h_1t+x)\varphi_3/h_1)}
{1-\exp(x\varphi_3/(2h_1))+\exp((h_1t+x)\varphi_3/h_1)},
\label{76}
\end{equation}
This exact solution shows how a solution of the problem with
a smooth initial condition of {\it double singularity} 
is developed during a finite time interval. 
A discontinuity arises at $t=1.4$. 
This discontinuity is of a special structure such that 
the left-hand discontinuity does not move, while 
the right-hand discontinuity moves  to the right.

Note that if the sign of the parameter $\varphi_3$ is changed
(i.e., we have  $\varphi_3=1$), 
then  Eq.~(\ref{1}) becomes an inverse parabolic equation.
However, the  solution (\ref{1}) exists and is bounded. 
Small perturbations of the initial condition relax in time.

{\it EXAMPLE 4}.
We consider the solution correspondign to the second mode for
the equation the fourth order of homogeneity. 
Suppose that the modified FhNS equation (\ref{1}) 
has the form
\begin{equation}
(1-u)u_t-(1-u)u_{xx}+
(-1/a_1+u/a_1-2a_1u\varphi_3)u_x-2(u_x)^{2}-
\varphi_3u(1-u^{2})=0.
\label{77}
\end{equation}
This equation has a solution of the form
\begin{equation}
u[x,t]=\frac{1-\exp(x/a1)}{1+\exp(x/a1)+\exp(x/a1-t\varphi_3)}.
\label{78}
\end{equation}
 
The solution shows how the structure develops 
from an initial state whose range is $[-0.5,1]$ 
to a traveling wave $u\in[0,1]$.
Note that it is possible to construct many solutions of such
type,  i.e., 
solutions that describe how a certain wave evolves from some
intial state. 

{\it EXAMPLE 5}.
We consider the modified
Fisher--Kolmogorov--Petrovskii--Peskunov  equation  
\begin{equation}
\label{79}
u_t-k_0u_{xx}-(-2k_0\varphi_{2}/h_2+h_2u)u_{xx}
-\varphi_{1}u+\varphi_{2}u^{2}=0.
\end{equation}

We consider an important example of  solutions 
to the Fisher--Kolmogorov--Petrovskii--Piskunov    
modified equation
and show that its solutions can be expressed 
in terms of higher transcendental functions 
such as Bessel functions or hypergeometric functions.
We also show how they are related to the B\"acklund transformation  
for the B\"urgers equation.

\begin{theor}
\label{th9}
Suppose $b_1=0$, $\varphi_3=0$,
\begin{equation}
h_1=-2k_0\varphi_2/h_2
\label{80}
\end{equation}
in equation~{\rm(\ref{3})}. 
The function $U(x,t)$ is a solution of
the linear parabolic equation 
\begin{equation}
U_t=r(x,t)U+k_0U_{xx},
\label{81}
\end{equation}
with the potential 
\begin{equation}
r(x,t)=\exp(2x/h_2+C_1(t)).
\label{82}
\end{equation}
The function $U(x,t)$ has the form
\begin{equation}
U (x, t) = v (z (x, t)),\qquad 
Z (x, t) = \exp \Big(x / (2h_2) + C_1 (t) /2)\Big) 
\label{83}
\end{equation}
and the function $v(z)$ is a solution of the equation
\begin{equation}
2h_2^{2}z\varphi_2v(z)+
h_2^{2}\varphi_1v'(z)+2k_0\varphi_2v'(z)+
2k_0z\varphi_2v''(z)
\label{84}
\end{equation} 
Then the function
$u(x,t)$ is a solution of Eq.~{\rm(\ref{79})} and has the form 
\begin{equation}
u(x,t)=(2k_0/h_2)\partial_{x}(\ln(U(x,t))),
\label{85}
\end{equation}
\end{theor}

{\it Proof}.
We substitute (\ref{85}) into the original equation (\ref{79}) 
and obtain a nonlinear homogeneous equation from which 
we exclude the derivatives $U_t$, $U_{xt}$ with the help of 
Eq.~(\ref{81}). 
Then, 
for the parameters specified in the assumptions of the theorem
by the change (\ref{84}),
we obtain ordinary defferential equations for higher
transcendental functions. 
The proof of Theorem~2.5 is complete.
$\Box$

It should be noted that the change~(\ref{85}) coincides in  form 
with the Cole--Hopf change of variables  
for the  B\"urgers equation.

The author is grateful to V.~G.~Danilov and S.~Yu.~Dobrohotov
for constant attention to his work and useful discussions
and to V.~P.~Maslov, and A.~D.~Polynin 
for constructive advice.



\end{document}